# High pressure flux growth, structural, and superconducting properties of *Ln*FeAsO (*Ln* = Pr, Nd, Sm) single crystals

N. D. Zhigadlo,[1,*] S. Weyeneth,[2] S. Katrych,[1,#] P. J. W. Moll,[1] K. Rogacki,[1,3] S. Bosma,[2] R. Puzniak,[2,4] J. Karpinski,[1,#] and B. Batlogg[1]

[1]*Laboratory for Solid State Physics, ETH Zurich, CH-8093 Zurich, Switzerland*
[2]*Physik-Institut der Universität Zürich, CH-8057 Zürich, Switzerland*
[3]*Institute of Low Temperature and Structure Research, PL-50-950 Wroclaw, Poland*
[4]*Institute of Physics, Polish Academy of Sciences, PL-02-668 Warsaw, Poland*

Single-crystals of the *Ln*FeAsO (*Ln*1111, *Ln* = Pr, Nd, and Sm) family with lateral dimensions up to 1 mm were grown from NaAs and KAs flux at high pressure. The crystals are of good structural quality and become superconducting when O is partially substituted by F (PrFeAsO$_{1-x}$F$_x$ and NdFeAsO$_{1-x}$F$_x$) or when Fe is substituted by Co (SmFe$_{1-x}$Co$_x$AsO). From magnetization measurements, we estimate the temperature dependence and anisotropy of the upper critical field and the critical current density of underdoped PrFeAsO$_{0.7}$F$_{0.3}$ crystal with $T_c \approx 25$ K. Single crystals of SmFe$_{1-x}$Co$_x$AsO with maximal $T_c$ up to 16.3 K for $x \approx 0.08$ were grown for the first time. From transport and magnetic measurements we estimate the critical fields and their anisotropy, and find these superconducting properties to be quite comparable to the ones in SmFeAsO$_{1-x}$F$_x$ with a much higher $T_c$ of $\approx 50$ K. The magnetically measured critical current densities are as high as $10^9$ A/m$^2$ at 2 K up to 7 T, with indication of the usual "fishtail" effect. The upper critical field estimated from resistivity measurements is anisotropic with slopes of ~ -8.7 T/K (*H* || *ab*-plane) and ~ -1.7 T/K (*H* || *c*-axis). This anisotropy (~ 5) is similar to that in other *Ln*1111 crystals with various higher $T_c$'s.

# I. INTRODUCTION

Further progress in exploring the physical properties of Fe-based superconductors and understanding the nature of superconductivity depends crucially on the availability of sufficiently large single crystals of high quality. However, as many groups have proved through trial and error [1-13], the crystal growth of *LnFePn*O (*Ln*1111, *Ln*: lanthanide, *Pn*: pnictogen) oxypnictides is a difficult task. A number of exciting features with respect to structural, electronic, magnetic, and superconducting properties are still unresolved or have not been adequately studied yet. Despite extensive experimental efforts over the past years, the *Ln*1111 single crystals grown by various methods are still limited in dimension, their superconducting critical temperature ($T_c$) is often reduced due to problems to maintain a certain stoichiometry of O and F, and some crystals contain inclusions of secondary phases or impurities, both affecting physical properties.

The application of flux as solvent in the crystal growth of oxypnictides developed significantly in the last few years. The original idea of flux usage is to initiate a reaction at a temperature much lower than normally used in solid state synthesis. This principle is occasionally employed at high pressure to enhance reaction rates. In the case of flux growth at ambient pressure in quartz ampoules, the growth temperature is limited and thus the solubility of starting components is very low and a long soaking time is required to grow even micrometer-sized crystals. In such conditions, difficulties arise due to the relatively high vapor pressures of arsenic (or phosphorous) and the vaporization loss of other components.

We have adopted a high-pressure and high-temperature (HPHT) technique and succeeded in the growth of *Ln*1111 single crystals using NaCl/KCl as a flux [2, 3]. This method has several advantages in comparison with the conventional ampoule method, since it avoids vaporization losses and allows control of the composition (doping) even at the high temperatures required for single crystal growth. After a systematic investigation of the parameters controlling the growth of *Ln*1111 crystals we identified the synthesis temperature and soaking time as the key parameters influencing the crystal size. However, even at optimal conditions the growth rate is extremely low and only crystals with linear sizes up to 300 μm are reproducibly grown [3]. Thus, to make single crystals of larger sizes, further improvements of the growth conditions are required.

Recent progress in the growth of millimeter-sized single crystals of pure and doped LaFeAsO and pure NdFeAsO using NaAs flux was reported by Yan *et al.* [11-13]. Although the size of the crystals was sufficiently increased, reported $T_c$'s were still limited due to difficulties in doping control. A later report [13] also revealed that some crystals show magnetic properties affected by ferromagnetic impurities due to trapped secondary phase inclusions. Nevertheless, these experiments suggest that NaAs flux has large solubility and diffusivity for oxygen, possibly due to formation of $NaAsO_2$. Motivated by these observations, we report in the current paper details of the first successful results of HPHT growth of millimeter-sized *Ln*1111 crystals, namely $NdFeAsO_{1-x}F_x$, $PrFeAsO_{1-x}F_x$, and $SmFe_{1-x}Co_xAsO$, using NaAs and KAs as flux. Their structural and superconducting properties are presented.

## II. EXPERIMENTAL DETAILS

For the growth of large size *Ln*1111 single crystals we applied the cubic anvil HPHT technique which was already developed earlier in our laboratory for growing superconducting $MgB_2$ crystals and various other compounds [14]. Powders of starting materials of *Ln*As, FeAs, $Fe_2O_3$, Fe, Co, and $LnF_3$ of high purity ($\geq 99.95$ %) were weighed according to the stoichiometric ratio, thoroughly grounded, and mixed with NaAs or KAs. Arsenide of NaAs and KAs were prepared by reacting Na and K metals with As pieces in evacuated and sealed ampoules at 600 °C for 12 h. The precursor to flux molar ratio was varied between 1:1 and 1:10. All work related to the sample preparation was performed in a glove box due to the toxicity of arsenic. A pellet containing precursor and flux was enclosed in a boron nitride (BN) container and placed inside a pyrophyllite cube with a graphite heater. After completing the crystal growth process, all remaining NaAs or KAs fluxes were dissolved in water. The structural properties of the single crystals were studied at room temperature on a *Bruker* X-ray single-crystal diffractometer. Data reduction and numerical absorption correction were performed using the *Bruker* AXS Inc. software package [15]. The crystal structure was determined by a direct method and refined on $F^2$, employing the SHELXS-97 and SHELXL-97 programs [16]. The elemental analysis of the crystals was performed by means of energy dispersive x-ray (EDX) spectrometry. Temperature-dependent magnetization measurements were carried out with a *Quantum Design* Magnetic Property Measurement System (MPMS) XL with the reciprocating sample option (RSO) installed. Four-point resistivity measurements were performed in a 14 Tesla *Quantum Design* Physical Property Measurement System (PPMS). Micrometer-sized platinum (Pt) leads were precisely deposited onto a plate-like crystal using a focused ion beam (FIB) method without altering the superconducting bulk properties [17].

**III. RESULTS AND DISCUSSION**

**A. Crystal growth**

Flux growth of *Ln*FeAsO oxypnictides is a technically complex task, requiring a combination of high pressure and high temperature and a careful control of the temporal evolution in time. We carried out a systematic study of *Ln*1111 crystal growth by the flux method using NaAs and KAs in order to understand the effects of parameters such as synthesis temperature, time, composition of the oxide mixture, amount of flux on the phase formation, and on the morphological properties of the crystals. Figure 1 provides the schematic illustration of the sample cell assembly, high-pressure synthetic process, and pictures of the obtained crystals. The sample cell assembly [Figs. 1(a) and 1(c)] is made of pyrophyllite cube with electrical leads made from discs of stainless steel. The graphite heater consists of a tube and two end caps. The boron nitride (BN) crucible of 6.8 mm or 8.0 mm in inner diameter and 10 mm in length fits inside the graphite tube and the sample of the starting materials is placed inside the BN crucible. The graphite furnace surrounds the sample and provides heat from the ends of the sample. The electrical current is brought to the furnace through stainless steel leads. Sample temperature is estimated by the predetermined relation between applied electrical power and measured temperature in the cell. The temperature calibration was conducted at the central, upper, and lower regions of the furnace. For the cell temperature of ~ 1450 °C, we estimated the temperature gradient across the sample to be around 70 °C. Such a thermal gradient in a high pressure cell is a result of heat transport by conduction. The furnace has a finite axial extent, inducing heat conduction along the assembly axis and causing a lower temperature away from the center of a sample in the axial direction. Thus the whole assembly produces a roughly parabolic temperature variation across the sample. The schematic view of the sample crucible adopted for high-pressure (HP) growth is shown in Fig. 1(c). The growth temperature corresponds to the temperature of the central position of the furnace where the BN crucible was fixed, as schematically illustrated in Fig. 1(c).

The assembled cell was compressed to 3 GPa at room temperature and the optimum growth conditions were tuned by varying the heating temperature, the reaction time, and the cooling rate. After optimization, we found the mixture of 1111 precursor with NaAs or KAs fluxes in the molar ratio less than 1:5 to be most effective for growing sizable $Ln$1111 crystals. Too high precursor-to-flux ratio prevents the $Ln$1111 phase formation and results in an increasing amount of impurity phases, which considerably affects the growth and appropriate doping. This is in contrast to the application of NaCl/KCl flux, where the precursor composition to flux ratio had very little effect on the phase formation. Thus, one can conclude that the 1111 phase formation is much easier in molten salts than that in NaAs and KAs. Therefore, it is not surprising that doping control and obtaining high $T_c$ superconductors through substituting of F for O is very difficult not only at ambient pressure [11-13] but also under HPHT crystal growth conditions. The synthesis temperature is the second very important parameter since once it exceeds 1450 °C, the 1111-type phase tends to decompose into various phases, such as $Ln$As, $FeAs_2$, $Ln$OF, FeAs, $Ln_2O_3$, *etc*. Thus to prevent the decomposition of the 1111 phase in the crystal growth process, the BN crucible was heated up to the maximum temperature of ~ 1350 – 1450 °C in 1 h, kept for 2 – 5 h, and cooled down with a cooling rate as shown in Fig. 1(b). We have tested two kinds of fluxes with low melting temperature ($T_m$), namely NaAs and KAs (for both $T_m$ ~ 800 °C [18]), expecting that their low $T_m$ and high solubility of oxygen allow stable synthesis of large $Ln$1111 single crystals. Application of both these fluxes led practically to the same results.

Growth features of the crystals using the NaAs and KAs solvents were studied at various growth temperatures for different starting compositions. Figures 1(c) and 1(d) show in a schematic way the conditions at the stages of crystal growth and the results of high pressure synthesis. Recovered samples were immersed into distilled water to dissolve the flux and then disaggregated by ultrasonic wave. As it seen in Fig. 1(c), the temperature gradient across the sample is an important parameter that influences the crystal growth process. Since the top and bottom parts of the BN crucible are coolest, crystallization starts at these points. As the furnace and subsequently the molten flux is gradually cooled, the melting point moves to the center, leaving a single crystal behind. As a consequence, most of the single crystals turned out to be in the range of ~ 400 – 700 μm in their lateral size, but some (at the bottom and top parts of BN crucible) reached sizes of ~ 1 mm, as shown in Fig. 1. In the crystal growth process the morphology is controlled by the growth rate anisotropy of crystal face and by the effect of molten flux on growth rate. Most of the crystals were found to exhibit plate-like shape with flat surfaces, because in the tetragonal structure of the 1111 phase, crystals with different faces grow with different rates, which are also accentuated by the presence of molten fluxes. It should be noted that the grown crystals do not seem to be in contact with the BN sample crucible, implying that the crystals were grown in the free space of the solvent area. By comparing of NaAs(KAs) flux growth with NaCl/KCl flux growth we conclude that the application of As-based fluxes is at least 3 times more efficient in obtaining bigger crystals, but it makes phase formation and doping control more complicated. In the growth process with all of these fluxes, contrary to ambient pressure synthesis, the kinetic barrier for the nucleation of the $Ln$1111 crystals may be smaller at HPHT growth, *i.e.* a large amount of spontaneous nucleation takes place when the growing temperature is high. It should be mentioned that spontaneous nucleation happening in a lot of places is typically unique for crystal growth under HPHT conditions [14, 19]. It is expected that large crystals would be obtained if the number of the nuclei could be suppressed in the spontaneous nucleation process. A further understanding of the current achievements and searching for a new solvent system, in which the kinetic barrier of nucleation is large, seems to be very important to establish a new route to grow large $Ln$1111 crystals with high $T_c$'s. On the other side, optimization of the growth condition with respect to the degree of supersaturation by controlling the temperature gradient is an important issue too.

**B. Superconducting transition temperature, crystal structure, and chemical composition**

Temperature dependences of the normalized magnetic moment, measured in a magnetic field of 0.5 mT parallel to the *c*-axis for the NdFeAsO$_{1-x}$F$_x$, PrFeAsO$_{1-x}$F$_x$ and SmFe$_{1-x}$Co$_x$AsO crystals are presented in Figs. 2(a) and 2(b). The effective superconducting transition temperature $T_c$ is defined as the cross point of the lines which are linearly extrapolated from the two regions of the high-temperature normal state and low-temperature superconducting state. Until now, for NdFeAsO$_{1-x}$F$_x$ and PrFeAsO$_{1-x}$F$_x$ we succeeded to obtain the crystals with highest $T_c$ of 39 K and 30 K, respectively. By comparing with the observed $T_c$ of optimized polycrystalline samples, we conclude that the doping level of our crystals is below optimal. Difficulties in doping control during the crystal growth and thus in obtaining crystals with high $T_c$ are mostly related to formation of NdOF and PrOF parasitic phases. Thus, the nominal/initial composition of the crystals studied does not agree with the real composition. In the case of substitution of Co for Fe, as we found for the SmFe$_{1-x}$Co$_x$AsO system, a doping of charge carriers is much simpler, since practically all nominally introduced Co ions substitute Fe. However, $T_c$ is much lower than in the case of the F substitution. As an example, the magnetization $M(T)$ results for selected crystals with various Co content are shown in Fig. 2(b). The superconducting transition temperature $T_c$ reaches a maximum of 16.3 K at the optimally doped level $x \approx 0.08$ [see Fig. 2(c)], which is higher than the value of 14.2 K reported in Ref. [20], but close to $T_c$ = 17.2 K as reported in Ref. [21]. The differences between the data presented here and the data published for polycrystalline samples may be associated with differences in sample preparation and uncertainties of concentration. We note that there is no appreciable difference in $T_c$ within one grown batch, which indicates macroscopic homogeneity of the crystals. The recorded magnetic response above $T_c$ is nearly zero, indicating that the samples do not contain magnetic impurities.

The single-crystal refinement data presented in Table 1 demonstrate the good structural quality of our flux-grown *Ln*1111 crystals. The evaluated details of the structure [see Table 1 and Fig. (3)] are consistent with the results of our previous x-ray diffraction studies [2, 3, 22-24]. The crystals belong to a layered ZrCuSiAs-type structure constructed by stacking the *Ln*O and FeAs layers, where the interlayer is connected by ionic bonding, and the intralayer is dominated by covalent bonding. As an example, Fig. 4 shows the reconstructed *0kl*, *h0l*, and *hk0* reciprocal space sections of the SmFe$_{0.92}$Co$_{0.08}$AsO single crystal measured at room temperature, where no additional phases (impurities, twins, or intergrowing crystals) were detected.

Compared to undoped SmFeAsO [$a$ = 3.9427(1) Å and $c$ = 8.4923(3) Å], both lattice parameters decrease to $a$ = 3.9410(2) Å and $c$ = 8.4675(7) Å for SmFe$_{0.92}$Co$_{0.08}$AsO. Though the $a$-axis does not change much, the $c$-axis shortens significantly with Co substitution, indicating that Co is indeed incorporated into the lattice. This is also a good indication that the superconductivity is not due to oxygen deficiency. It was previously reported that oxygen deficient SmFeAsO$_{1-y}$ samples showed a noticeable decrease in the $a$-axis (about 0.9 %) [25], an effect not seen in the crystals studied here. Further confirmation of successful substitution is also clearly seen if one compares the Fe-As distances. For SmFeAsO, as well as for many others $Ln$1111 systems, the Fe-As distance is fixed at ~ 2.40 Å, while for the SmFe$_{0.92}$Co$_{0.08}$AsO, it decreases to 2.3879(3) Å. A similar trend in shrinkage of the $c$-axis and thus the cell volume was observed for the SmFe$_{1-x}$Co$_x$AsO, LaFe$_{1-x}$Co$_x$AsO, NdFe$_{1-x}$Co$_x$AsO, and CeFe$_{1-x}$Co$_x$AsO polycrystalline samples [20, 26-28]. This can be interpreted as a result of an increase of the density of negative charge in the FeAs layers induced by the Co doping, which leads to the strengthening of the interlayer Coulomb attraction. It is also interesting to note that the threshold and optimal electron doping level (~ 0.08 electron/Fe) is close to that of $Ln$FeAsO$_{1-x}$F$_x$, notwithstanding the different doped layer.

The resulting stoichiometries of the $Ln$1111 crystals were revealed by x-ray structure refinement and were further confirmed by energy dispersive x-ray spectroscopy (EDX) analysis. Both methods show that the ratio of lanthanide, iron, and arsenic is equal to 1:1:1, consistent with the nominal composition. The light elements of oxygen and fluorine can not be detected accurately; therefore, we could not determine the actual doping level of NdFeAsO$_{1-x}$F$_x$ and PrFeAsO$_{1-x}$F$_x$ crystals. Because of the almost equal number of electrons, both Co and Fe atoms were considered to be identical in the x-ray refinement. For all Co substitutions from 4 to 13 at% the real Co content determined by EDX is in good agreement with the nominal one. The actual composition of the single crystals was taken as the average of 5 spots measured on the crystal. In the following, the average experimentally determined $x$ values will be used to identify all the crystals rather than the nominal concentration. The compositional spread over a wide area on the sample surface for each concentration is less than 0.002, demonstrating relative homogeneity of the Co substituted single-crystal samples. The optimal Co concentration for superconductivity in SmFe$_{1-x}$Co$_x$AsO was determined to be at $x \approx 0.08$. These results are in stark contrast to the nonmonotonic and scattered results found for the Ca(Fe$_{1-x}$Co$_x$)$_2$As$_2$ crystals grown from Sn flux, for which solubility problems in Sn causes transition broadening and make measurements on homogeneous samples difficult [29].

## C. Magnetic properties of PrFeAsO$_{0.7}$F$_{0.3}$ single crystals

For the case of the *Ln*1111 family of Fe-based superconductors, most of the data concerning the distinct temperature and orientational dependences of $H_{c2}$ and $J_c$ were obtained on the crystals with $T_c$ ranges between 35 and 50 K. Limited amount of such data are available for the *Ln*1111 crystals with $T_c$ lower than 30 K. Usually, for underdoped *Ln*1111 crystals, the width of the superconducting transition broadens substantially due to the difficulty to maintain homogeneous doping of F for O, which makes the measurements and analysis of $H_{c2}$ and $J_c$ more complicated. As we have shown in the previous section, the application of NaAs and KAs fluxes allows the growth of underdoped *Ln*1111 crystals with low $T_c$, but relatively sharp transitions. The availability of such crystals opens the possibility for further exploration of their superconducting properties.

The PrFeAsO$_{0.7}$F$_{0.3}$ single crystals have been preliminary characterized by measurements of the magnetic moment *m* as a function of temperature in an applied dc field of 0.2 or 0.5 mT. The transition to the normal state, obtained upon heating from the zero-field-cooled superconducting state, has a width of ~ 3 K, so is relatively sharp, as for underdoped crystals with $T_c \approx 25$ K. For the upper critical field and critical current studies, a large crystal with a mass of 0.55 mg and dimensions of 1.00 × 0.63 × 0.13 mm$^3$ was selected. The density of this crystal was estimated be 6.6 Mg/m$^3$, which is close to the x-ray density, 7.2 Mg/m$^3$. The upper critical field was obtained from the temperature dependence of the magnetic moment measured in dc magnetic field up to 7 T. For *H* parallel to the *ab*-plane, the data obtained at fields higher than 6 T were skipped due to the small signal to noise ratio. Figure 5 shows the temperature dependence of the upper critical field, measured in *H* parallel to the *c*-axis, $H_{c2}^{\|c}$, and to the *ab*-plane, $H_{c2}^{\|ab}$, together with the upper critical field anisotropy, $\gamma_H = H_{c2}^{\|ab} / H_{c2}^{\|c}$. This anisotropy is surprisingly low at $T_c$ ($\gamma_H \approx 1$) and increases to about 6 with decreasing temperature. The temperature dependence of $\gamma_H$ is similar to that observed for the almost optimally doped Nd1111 single crystals [30] with $T_c \approx 47$ K, and is opposite to that obtained for the SmFe$_{0.92}$Co$_{0.08}$AsO crystal with $T_c \approx 16$ K (see discussion of transport properties of SmFe$_{0.92}$Co$_{0.08}$AsO). However, it is a rather intricate task to compare the temperature dependences of anisotropies obtained with different criteria, even if the same experimental technique is employed, which was not the case here. Importantly, the anisotropy of PrFeAsO$_{0.7}$F$_{0.3}$ crystals, especially just below $T_c$, is rather sensitive to the chosen criteria (see for example Fig. 15 in Ref. [30]). Besides, crystals grown by the NaAs and KAs flux method are much thicker compared to those grown from NaCl/KCl flux; thus the possibility for occurrence of planar defects parallel to the *ab*-plane is much higher. This may broaden the transition at $T_c$ and affect the curvature of $H_{c2}(T)$ near $T_c$, which could mask two-band effects.

The shielding effect of the PrFeAsO$_{0.7}$F$_{0.3}$ single crystal was evaluated from the dc magnetic moment measured versus magnetic field at 5 K, for the crystal in a virgin zero-field-cooled state (see the inset of Fig. 6). The demagnetizing factor was determined by approximating the crystal shape with an ellipsoid. For the magnetic moment $m(H)$ measurements, the ratio $M/H_{int}$ was estimated in the Meissner state, where $M = m/V$ is the volume magnetization and $H_{int}$ is the magnetic field corrected for demagnetizing effects. The ratio $-M/H_{int}$ was close to 0.9 for both field orientations, $H \parallel c$-axis and $H \parallel ab$-plane. Comparing with the ideal value of 1, this result indicates almost perfect shielding. The most probable reason for the small difference between the ideal and the evaluated $M/H$ values would be irregular shape of the crystal.

Magnetic moment hysteresis loops, $m(H)$, were measured at constant temperatures, for $H$ oriented parallel to the *c*-axis or to the *ab*-plane, in order to determine the persistent critical current density in the *ab*-plane, $j_c^{ab}$, or along the *c*-axis, $j_c^c$, respectively. Figure 6 shows the $m(H)$ curves obtained at 5, 10 and 15 K for the PrFeAsO$_{0.7}$F$_{0.3}$ single crystal ($T_c = 25.2$ K), for $H$ oriented parallel to the *c*-axis. No ferromagnetic contribution to $m(H)$ was observed, contrary to results reported for Sm1111 polycrystalline samples with $T_c = 46$ K [31]. The only small broadening of the magnetization loops with increasing $H$, the so-called fishtail effect, was detected at low temperatures and low fields. This suggests a relatively weak pinning and low critical current density in higher magnetic fields. Similar magnetization loops, however with larger paramagnetic background, have been obtained for a configuration $H$ parallel to the *ab*-plane. The results allow us to estimate the critical current density anisotropy, which is an important parameter for practical applications.

The Bean critical-state model [32] with corrections for demagnetizing effects was applied to obtain the critical current densities from the magnetization loops. The calculation of $j_c^{ab}(H)$ was performed by using the formula $j_c^{ab}(H) = 2\Delta M(H)/b(1-b/3a)$, where $j_c^{ab}$ is in A/m$^2$, $\Delta M(H)$ is the width of the $M(H)$ loop in A/m, and $a$ and $b$ are the crystal length and width in m, respectively. The $\Delta M(H)$ values at fields above the first magnetization peak (see Fig. 6) were used to make the Bean model applicable. For the calculation of $j_c^c(H)$ the simplified Bean formula was used: $j_c^c(H) = 3\Delta M(H)/w$, where $w$ is a scaling factor taken as a half of the averaged dimensions of the crystal side perpendicular to the field. Figure 7 shows the $j_c^{ab}(H)$ and $j_c^c(H)$ dependences at several temperatures. Relatively low values of $j_c$'s and strong field and temperature dependences obtained for $j_c^{ab}(H)$ reveal rather weak pinning of vortices for $H \parallel c$-axis. Thus any small volume disorders and other isotropic pinning centers do not work effectively in this crystal. On the other hand, $j_c^c(H)$ obtained for $H \parallel ab$-plane is much more resistant to an increase of both field and temperature and even exceeds $j_c^{ab}(H)$ at higher fields. This seems to be a consequence of the layered structure of PrFeAsO$_{0.7}$F$_{0.3}$ and a result of intrinsic pinning by PrO layers. For higher fields, $j_c^c(H)$ is more than 10 times larger that $j_c^{ab}(H)$. At low temperatures a weak fishtail effect appears for the $m(H)$ loops resulting in small maxima in both $j_c^c(H)$ and $j_c^{ab}(H)$ dependences.

We note that $J_c$ evaluation of anisotropic high-$T_c$ single crystals is not trivial for $H$ parallel to the *ab*-plane. We took into account the sample aspect ratio and we are aware of the well known formulas derived by Gyorgy *et al.* [33], which are particularly appropriate for materials with relatively large anisotropy. The compounds we have studied generally have low anisotropy (~ 5), therefore, we used properly the simplified formulas. Using the same formulas for YBaCuO crystals with relatively high anisotropy (~ 20), Dinger *et al.* [34] obtained similar results as derived by Gyorgy *et al.* [33] and thus show the approximation to introduce minor changes in the number. Therefore, the drop of $J_c$ anisotropy, as mentioned above, can not be changed qualitatively by applying a different formula (what we actually tested) and thus remains as the general tendency of the $J_c$ behavior of underdoped Pr1111 in magnetic fields. It is widely known for high-$T_c$ materials, both cuprates and pnictides (many examples in literature), that the variations of the carrier concentration (e.g. by substitution) not only influences the critical temperature, but also the over-all electronic properties. In particular, also the pinning landscape and thus the critical current densities are affected (see, for example, Ref [35]). We are mindful of the fact that the anisotropy of $J_c$ is not an intrinsic and generic property of a superconductor, and therefore various types of pinning centers may be introduced by the various ways of substitution, and/or growth method. Furthemore, one would have to specify in great detail the criteria used to define $J_c$ in different materials measured by different techniques.

Here we note a few particular details. At low temperatures and low fields, the anisotropy of $J_c$ we observe in PrFeAsO$_{0.7}$F$_{0.3}$ is about 3 and this anisotropy first slightly increases and then decreases with growing field. The anisotropy becomes less than 1 at fields higher than 6 T due to the fishtail effect which develops for $H$ oriented parallel to the *ab*-plane. This result is qualitatively different from that we observed for optimally doped Sm1111 single crystals with $T_c \sim 50$ K [17], where a fishtail effect was observed for $H$ perpendicular to the *ab*-plane. Thus, we show here the real change of the pinning mechanism in underdoped Pr1111 and this is one of a several new messages in the paper. The pinning mechanism in the *Ln*1111 superconductors is still not clear; it is complicated by material inhomogeneity on mesoscopic and macroscopic length scales. Nevertheless, a growing number of experimental data suggests that the major contribution comes from collective pinning of vortex lines by microscopic defects by the mean-free-path fluctuation mechanism [35], which means that dopant atoms play an important role in vortex pinning and quasiparticle scattering. Contrasting behavior of optimally doped high-$T_c$ Sm1111 and underdoped low-$T_c$ Pr1111 compounds implies the dependence of $J_c$ on $T_c$ and on the doping level; the more dopants we added, the higher $T_c$ but also the stronger the pinning. From the defect-vortex interaction, one might hope to extract information on electronic scattering mechanism in the FeAs based superconductors, however, much more work should be done to further clarify this point. As a first step in this direction, recently, we have discovered in SmFeAs(O,F) crystals a distinct change in the nature of the vortices from pinned Abrikosov-like to mobile Josephson-like vortices as the system is cooled below its critical temperature [36].

**D. Magnetic and transport properties of SmFe$_{0.92}$Co$_{0.08}$AsO single crystals**

To the best of our knowledge, crystals of SmFe$_{1-x}$Co$_x$AsO with $T_c$ as high as 16.3 K were grown for the first time; therefore, their properties were studied in detail. The temperature dependence of the zero-field cooled (zfc) magnetic moment of a single crystal of nominal composition SmFe$_{0.92}$Co$_{0.08}$AsO with approximate dimensions of 0.40 × 0.40 × 0.08 mm$^3$ in 1 mT magnetic field applied along its *ab*-plane is shown in Fig. 8. The crystal exhibits a sharp transition to the superconducting state at $T_c$ = 16.3 K. The strong diamagnetic signal below $T_c$ is consistent with bulk superconductivity.

The field dependence of the magnetic moment was studied for external magnetic fields $H_{ext}$ applied along the *c*-axis and along the *ab*-plane for various temperatures below $T_c$. These allowed us to gain information on the first penetration field $H_p$, which denotes the magnetic field above which vortices enter the sample, and the related lower critical field $H_{c1}$. Since in the Meissner state the susceptibility of the studied superconductor is -1, demagnetization effects of the superconducting sample are crucial and must be taken into account. We estimate from the crystal dimensions the approximate demagnetization factors $N_c \approx 0.8$ and $N_{ab} \approx 0.1$ along the *c*-axis and in the *ab*-plane, respectively. The internal magnetic field $H_{int}$ depends accordingly on the crystallographic direction along which the magnetic field was applied, *i.e.*, for the superconductor in Meissner state $H_{int}(1-N_i) = H_{ext}$.

Figures 9(a) and 9(b) present the as measured magnetic moment in low external magnetic fields applied along the *c*-axis and along the *ab*-plane, respectively. The magnetic moment $m(H_{ext})$ is linear as a function of magnetic field in the Meissner state and shows an upwards curvature above $H_p$ due to the entrance of vortices into the bulk of the crystal. In Figs. 10(a) and 10(b) the quantity $(BV)^{1/2}$ is plotted as a function of the internal magnetic field. Here, $B$ denotes the magnetic induction and $V$ is the sample volume, $V \approx 0.013$ mm$^3$. Since $B = \mu_0(M + H_{int}) = \mu_0(m/V + H_{int}) = 0$ in the Meissner state, it is possible to estimate the field $H_{c1}$. Since the quantity $BV$ empirically scales as the square of $H_{int}$ above $H_{c1}$ [23, 37], a plot of $(BV)^{1/2}$ as a function of $H_{int}$ allows a straightforward determination of $H_{c1}$. The sudden increase from zero occurs due to the penetration of vortices at $H_p$. In Fig. 10(c), the temperature evolution for $H_{c1}$ for the studied SmFe$_{0.92}$Co$_{0.08}$AsO single crystal is shown. The zero temperature estimates $\mu_0 H_{c1}^{\|c}(0) \approx 11$ mT and $\mu_0 H_{c1}^{\|ab}(0) \approx 4$ mT are made. The $H_{c1}$- anisotropy at low temperatures is estimated to be ~ 2.8, which is in good agreement with previous reports [38, 39]. Since according to [40]

$$H_{c1}^{\|c} = \frac{\Phi_0}{8\pi\mu_0\lambda_{ab}^2}\left[2\ln\left(\frac{\lambda_{ab}}{\xi_{ab}}\right) + 1\right],$$

$$H_{c1}^{\|ab} = \frac{\Phi_0}{8\pi\mu_0\lambda_{ab}\lambda_c}\left[\ln\left(\frac{\lambda_{ab}\lambda_c}{\xi_{ab}\xi_c}\right) + 1\right]. \quad (1)$$

it is possible to estimate the penetration depths being equal to $\lambda_{ab} \approx 250$ nm and $\lambda_c \approx 970$ nm, invoking the coherence lengths of $\xi_{ab} \approx 4$ nm and $\xi_c \approx 0.8$ nm estimated by the resistivity measurements discussed below.

From the field dependent measurements $m(H)$ in magnetic fields up to 7 T it is possible to extract information on the critical current density. In Figs. 11(a) and 11(c), the irreversible $m(H)$ is shown, measured for both increasing [$m_\uparrow(H)$] and decreasing [$m_\downarrow(H)$] magnetic fields. The critical current density $j_c^c$ and $j_c^{ab}$ is derived from the difference of $m_\downarrow(H)$ and $m_\uparrow(H)$ [see Figs. 11(b) and 11(d)]. Very similar values exceeding $10^9$ A/m$^2$ are derived for both magnetic field configurations. When the field is applied parallel to the *c*-axis the slight increase in critical density for higher magnetic fields is observed and it may indicate an increase in the effectiveness of pinning centers giving rise to a "peak effect". A similar behavior was found in F and Th-substituted SmFeAsO single crystals [17, 22]. Interestingly, when the field is applied parallel to *ab*-plane the critical current became practically field independent.

To determine $H_{c2}$ we studied the temperature dependence of the resistance for a SmFe$_{0.92}$Co$_{0.08}$AsO single crystal with the field applied parallel to the (Fe/Co$_2$As$_2$) layers ($H \parallel ab$) and perpendicular to them ($H \parallel c$), in various magnetic fields (0 – 14 T, 1 T step for the field in the *ab*-plane and 0.5, 1 or 2 T steps for the field along the *c*-axis) [see Figs. 12(a) and 12(b)]. While magnetic field of 14 T applied in the *ab*-plane suppress $T_c$ by approximately 3 K, the superconducting transition is seen to be shifted considerably to lower temperatures for the fields applied along the *c*-axis. Similar behavior was reported for Sm1111 substituted with F for O or with Th for Sm [17, 22, 23]. Interestingly, present magnetoresistance data of $\rho(T, H)$ for SmFe$_{0.92}$Co$_{0.08}$AsO crystal show remarkably different behavior to that recently reported for LaFe$_{0.92}$Co$_{0.08}$AsO crystal [41]. The latter case reminds $\rho(T, H)$ behavior observed for SmFeAs$_{0.5}$P$_{0.5}$O$_{0.85}$ crystal [23], where parallel shift of resistive transition curves was observed for the fields applied along all crystallographic axes, as usually seen in conventional superconductors. In order to clarify the origin of such a contrast, more systematic experimental and theoretical work is required. The inset in Fig. 12(a) shows a typical cool-down resistivity curve for SmFe$_{0.92}$Co$_{0.08}$AsO single crystal in the temperature range 2 – 300 K for current flowing within the planes in absence of an externally applied field. Similar to LaFe$_{1-x}$Co$_x$AsO, PrFe$_{1-x}$Co$_x$AsO, and SmFe$_{1-x}$Co$_x$AsO polycrystalline samples [42-44], in optimally doped SmFe$_{1-x}$Co$_x$AsO single crystals there is a crossover from metal to insulator as the temperature decreases. The normal state displays metallic conduction at high temperature, but it changes into semiconducting-like before reaching the superconducting transition, making the Co-doped superconductors different from the F-doped ones. However, it should be noted that electrons are directly doped into FeAs layers by partial replacement of Fe with Co, whereas the F substitution is regarded as indirect electron doping through SmO layers. Thus, the replacement of Fe introduces scattering centers in the superconducting layers. This is one of the reasons for the suppression of $T_c$ [while the $h_{Pn}$ = 1.3489(9) Å (see Table 1) favor high $T_c$] and the resistivity upturn in lower temperatures. It is evident that the resistance data do not show a drop in resistivity at 140 K, which is consistent with previous reports on Co-doping in the $Ln$1111 system [27, 42-44]. Once superconductivity appears, the indication of a magnetic transition seems to disappear.

Figure 13 shows the phase diagram obtained from magnetization and resistivity measurements. The irreversibility field $H_{irr}$, estimated from the onset of irreversible magnetization in the SQUID measurements, is drawn for both configurations, $H$ parallel to the $c$-axis and $H$ parallel to the $ab$-plane and compared with $H_{zero\ res.}$ estimated at the temperature where the onset of zero resistivity in the $\rho(T)$ recorded in magnetic field $H$ was found noticeable. The upper critical field $H_{c2}$, defined as the magnetic field where linearly extrapolated normal state resistivity $\rho_n$ is suppressed by 50%, was determined. The upper critical field slopes $\mu_0 dH_{c2}/dT \sim$ -8.7 T/K for $H \parallel ab$ and $\sim$ -1.7 T/K for $H \parallel c$ were determined using the linear part of $H_{c2}(T)$. These slopes suggest very high values of $H_{c2}(0)$. Furthermore, in both directions, we find that $H_{c2}(T)$ dependence is almost linear, with no sign of saturation. Nevertheless, the slope in $H_{c2}(T)$ for $H \parallel c$ is much lower than the one observed in Sm1111 doped with F for O and Th for Sm [22, 23]. This may be correlated to the strong disorder induced by the substitution of Fe by Co in the conducting layer.

The ratio $H_{c2}^{\parallel ab}/H_{c2}^{\parallel c}$ provides a rough estimation of the upper critical field anisotropy $\gamma_H$ for the temperatures below $T_c$. In the inset of Fig. 13 it is shown that the $\gamma_H$ value of SmFe$_{0.92}$Co$_{0.08}$AsO is ~ 8 near $T_c$ and then decreases to ~ 5 with decreasing temperature. This value of $\gamma_H$ and its temperature dependence is similar to that of other Fe-based superconductors [30, 45, 46]. Comparing the $H_{c2}$-anisotropy of the crystals grown with NaAs (KAs) flux with that of other $Ln$1111 crystals in previous studies [3, 30, 38, 39, 45-48], we note an interesting similarity: for crystals grown from various fluxes, with various lanthanides, and with $T_c$ ranging from 16 K to ~ 50 K, the anisotropy is typically around 5 (±1) (in the immediate vicinity to $T_c$ this value depends on the criteria and method of defining "$H_{c2}$"). Close to $T_c$ the $\gamma_H$ of the 1111-type phase is bigger than that of the 122-type [49-52] and is comparable with that of the 10-4-8 phase [53]. Based on the results presented here and on the results of other studies, this temperature dependence of $\gamma_H$ and a small anisotropy at $T \ll T_c$ seem to be a general feature of the Fe-based superconductors and can be considered as one of the arguments in support of the common multi-band scenario proposed for FeAs-based superconductors [54].

There are still extended debates in the literature on the role of Co in inducing superconductivity in Fe-based pnictides [55-59]. Naturally, the Co substituted systems have been widely referred to as electron-doped. However, this assumption is still not verified experimentally. Alternatively, it has been argued that Co is isovalent to Fe and that the main role of the Fe(Co) substitution is to introduce a random impurity potential [55]. Whether or not Co substitution is able to charge dope the Fe ions is a major issue that may help to identify the specific mechanism of superconductivity in Fe-based pnictides. For the case of Co substituted BaFe$_2$As$_2$, the lack of a Fe K edge absorption shift implies that Co is not charge doping the Fe ions, which are thought to be responsible for the superconductivity [57]. Rather it was argued that superconductivity may emerge due to bonding modifications induced by the substitute atom that weakens the spin-density-wave ground state by reducing the Fe local moments. The most relevant structural parameter is the Fe-As bond distance, since it controls directly the chemical pressure on Fe. Our results indicate that in the SmFe$_{0.92}$Co$_{0.08}$AsO crystal the (Fe,Co)-As bond distance is slightly [~ 0.012 Å (0.5%)] reduced with respect to the pure compound. Thus the contraction of the Fe-As bond length reported here, as well as the application of pressure or substitution of As by P [23], produce at least one common qualitative trend: a reduction of the local Fe magnetic moments. It was also shown that diluting the Fe plane by Ni, Rh, and Ir atoms again destroys the magnetic order and induces superconductivity [60].

## IV. CONCLUSIONS

We adopted the high pressure crystal growth method and carried out a systematic investigation of the parameters controlling the growth of $Ln$1111 crystals, including the thermodynamic variables, reagent composition and kinetic factors such as reaction time and cooling rate. The high-pressure NaAs and KAs flux-growth technique presented here yields millimeter-sized $Ln$1111 crystals that can be readily separated and studied. In comparison with NaCl/KCl flux-growth these fluxes are at least three times more efficient in obtaining large sized crystals. However, the 1111 phase formation and chemical composition are more difficult to control. X-ray structural investigations confirmed good structural quality of the crystals, and show modifications due to substitutions, which are linked to superconducting properties. The sufficiently large size of the crystals makes possible the development of a whole suite of single-crystal experimental techniques not previously possible for $Ln$1111-type pnictides.

The so-called fishtail effect was detected at low temperatures and low fields in underdoped PrFeAsO$_{0.7}$F$_{0.3}$ crystals with $T_c$ = 25.2 K. This suggests a relatively weak pinning and low critical current density in higher fields.

Magnetic measurements performed on the SmFe$_{0.92}$Co$_{0.08}$AsO crystal with a $T_c$ = 16.3 K show a relatively high critical current density of 10$^9$ A/m$^2$ at 2 K almost independent of the magnetic field. The upper critical field $\mu_0H_{c2}$ in SmFe$_{0.92}$Co$_{0.08}$AsO extracted from the resistivity measurements is anisotropic with slopes of ~ -8.7 T/K ($H \parallel ab$-plane) and ~ -1.7 T/K ($H \parallel c$-axis), sufficiently far below $T_c$. The upper critical field anisotropy $\gamma_H$ is temperature dependent, as already reported for other NdFeAsO$_{1-x}$F$_x$, SmFeAsO$_{1-x}$F$_x$, and Sm$_{1-x}$Th$_x$FeAsO compounds. This unusual temperature behavior of $\gamma_H$ observed in Co-substituted

SmFeAsO further supports the common multi-band scenario proposed for FeAs-based superconductors.

**Acknowledgments**

We would like to thank P. Wägli for the EDX analysis. This work was supported by the Swiss National Science Foundation, the National Center of Competence in Research MaNEP (Materials with Novel Electronic Properties), and by the National Science Centre of Poland based on decision No. DEC-2011/01/B/ST3/02374. S.K. and J.K. acknowledge support from ERC project Super Iron.

∗ zhigadlo@phys.ethz.ch

# Present address: Institut de Physique de la Matière Complexe, École Polytechnique Fédérale de Lausanne (EPFL), CH-1015 Lausanne, Switzerland

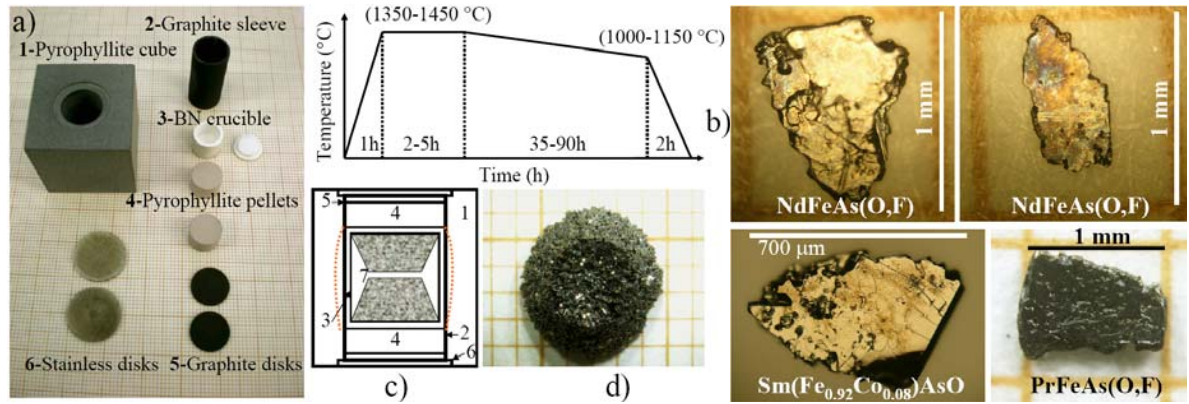

**FIG. 1.** (Color online) Schematic illustration of the sample cell assembly, high-pressure synthetic process and examples of crystals. (a) The cell assembly for crystal growth: 1) Pyrophyllite cube, 2) Graphite sleeve, 3) BN sample crucible, 4) Pyrophyllite pellets, 5) Graphite disks, 6) Stainless disks. (b) Typical temperature-time profile of the single crystal growth. Pieces with collection of crystals [labeled 7 in (c) and photo in (d)] are found at the top and bottom parts of the crucible after dissolving the rest of NaAs or KAs fluxes in water. Dashed line in Fig. 1(b) indicates the temperature gradient in the high-pressure cell assembly. $Ln$1111 crystals are shown on the right side of the figure.

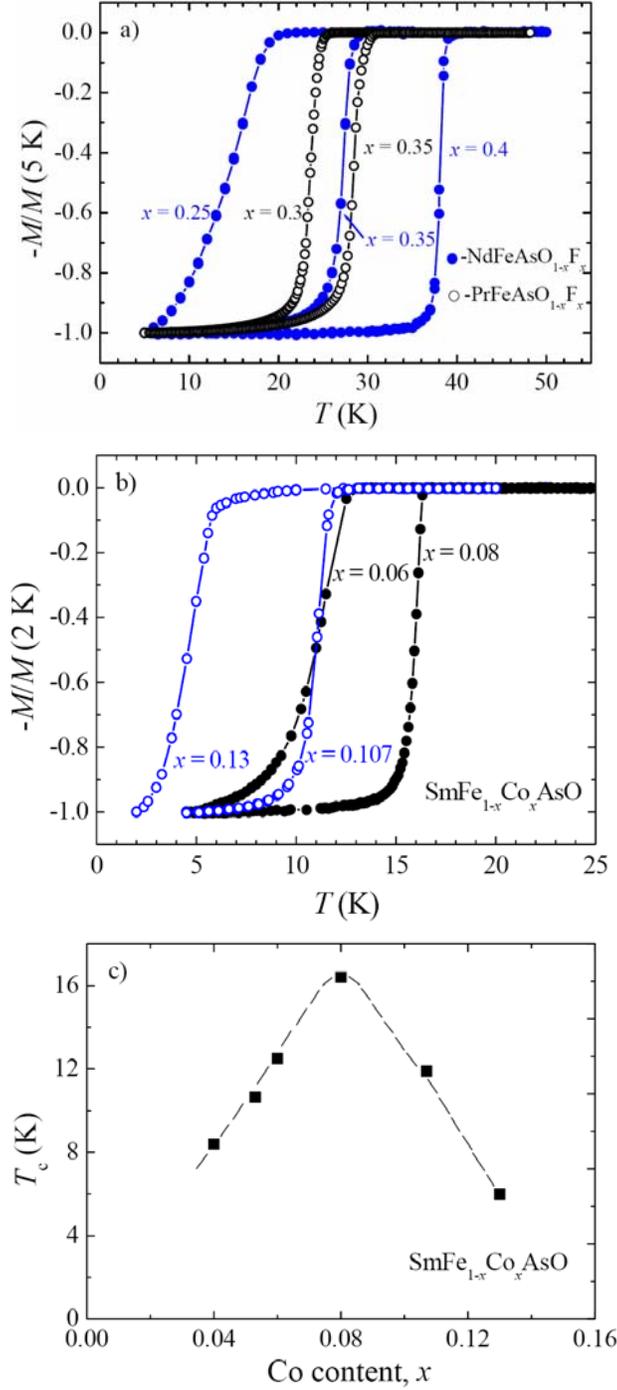

**FIG. 2.** (Color online) Normalized magnetic moment vs. temperature for NdFeAsO$_{1-x}$F$_x$ (a), PrFeAsO$_{1-x}$F$_x$ (a), and SmFe$_{1-x}$Co$_x$AsO (b) single crystals. The measurements were performed in a field of 0.5 mT, after cooling in a zero field with $H \parallel c$-axis. (c) $T_c$ vs. Co content $x$ (determined with EDX) for SmFe$_{1-x}$Co$_x$AsO single crystals.

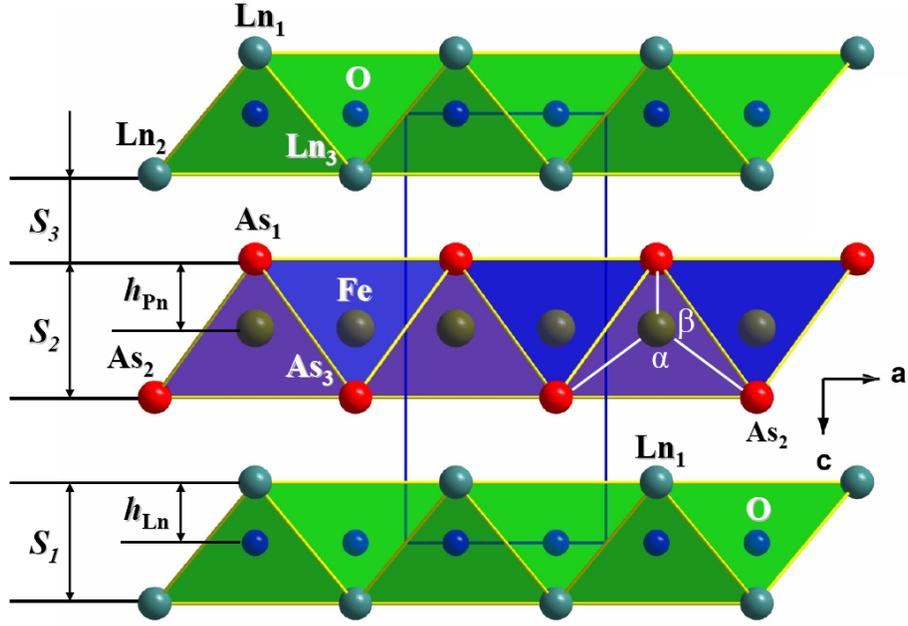

**FIG. 3.** (Color online) Schematic representation of the projection of the *LnFePnO* (*Ln*-1111, *Ln*: lanthanide, *Pn*: pnictogen) lattice on the *ac*-plane (for details see Table 1). *Ln*O and Fe*Pn* layers are stacked alternately. $h_{Pn}$ is the height of the As/P atoms above the plane of iron atoms. $h_{Ln}$ is the height of the rare-earth metal atom above the plane of oxygen atoms. $S_1$ is the thickness of a charge supplier layer. $S_2$ is the thickness of conducting layer. $S_3$ is the interlayer distance. The As-Fe-As bond angles α and β indicates the deviation from a regular FeAs$_4$ tetrahedron in which α and β are equal to 109.47°.

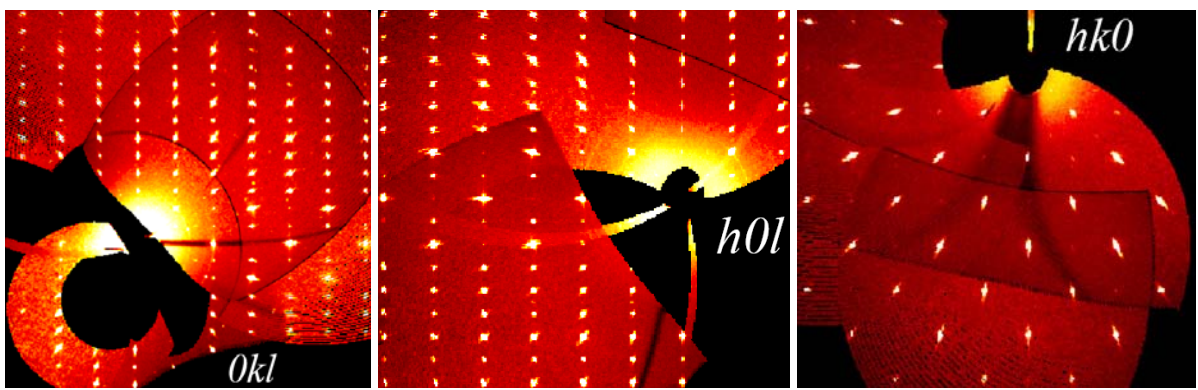

**FIG. 4.** (Color online) The reconstructed *0kl*, *h0l*, and *hk0* reciprocal space sections of the single crystal SmFe$_{0.92}$Co$_{0.08}$AsO.

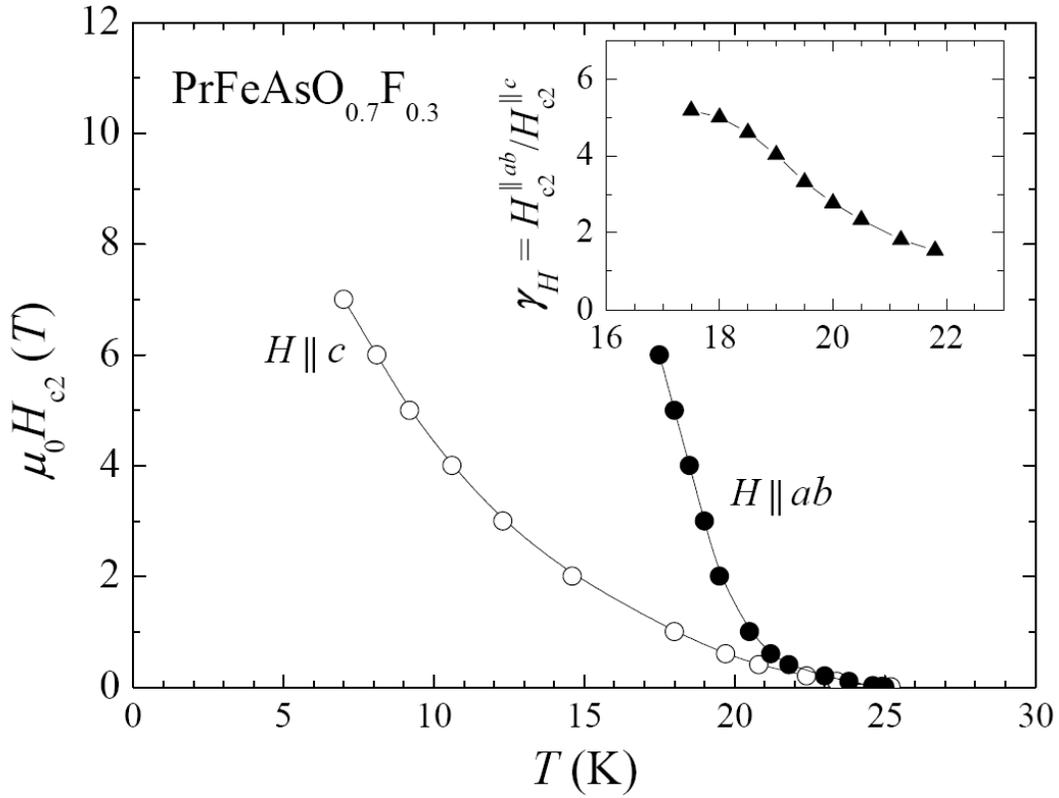

**FIG. 5.** Upper critical field of the underdoped PrFeAsO$_{0.7}$F$_{0.3}$ single crystal ($T_c$ = 25.2 K), for $H$ oriented parallel ($H \parallel ab$) and perpendicular ($H \parallel c$) to the $ab$-plane. The $H_{c2}(T)$ results have been derived from the $M(T)$ data obtained at constant dc field. The inset shows the upper critical field anisotropy.

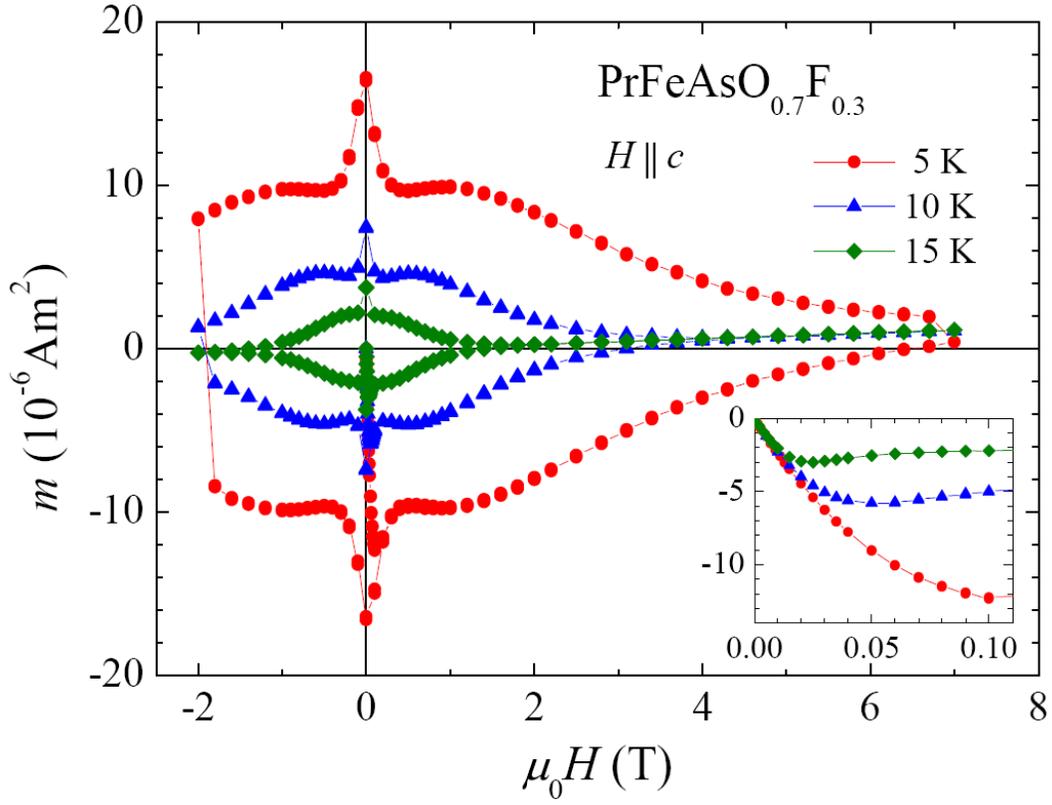

**FIG. 6.** (Color online) Magnetic moment loops of the underdoped PrFeAsO$_{0.7}$F$_{0.3}$ single crystal ($T_c$ = 25.2 K), for $H$ oriented parallel to the $c$-axis. The inset shows the initial part of the loops obtained in the virgin state. The initial slope of these curves was used to evaluate the shielding susceptibility and estimate the superconducting volume fraction (see text for details).

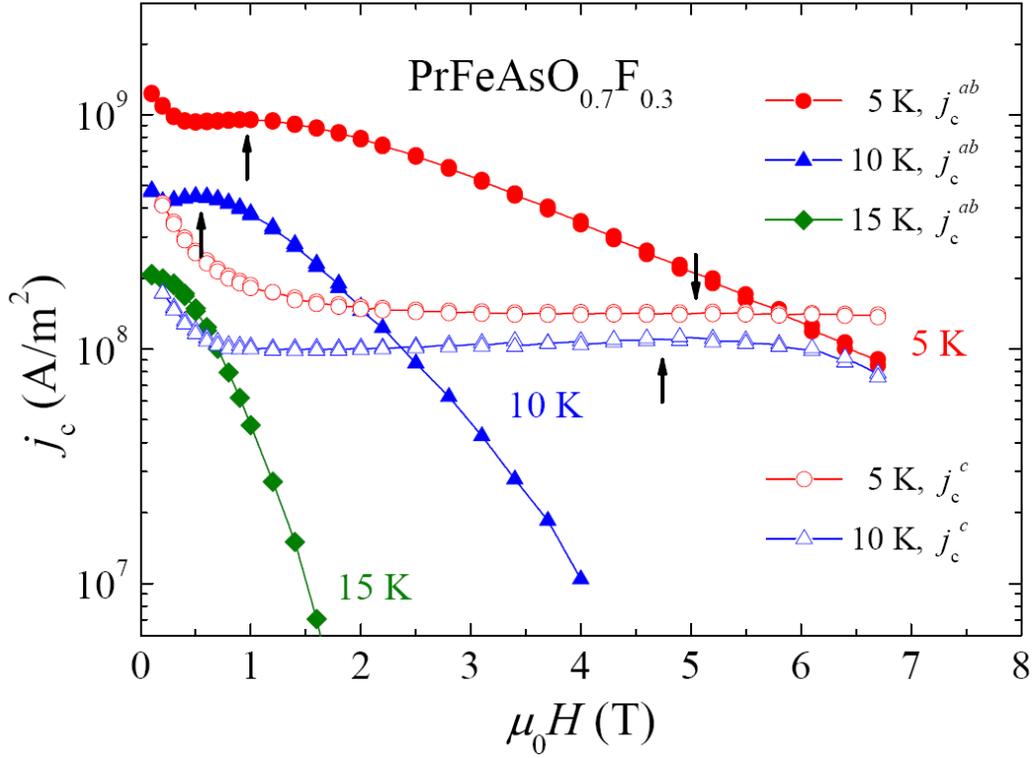

**FIG. 7.** (Color online) Critical current density in the *ab*-plane, $j_c^{ab}$ (closed symbols), and perpendicular to the plane, $j_c^c$ (open symbols), versus magnetic field for the underdoped PrFeAsO$_{0.7}$F$_{0.3}$ single crystal ($T_c$ = 25.2 K). The $j_c(H)$ results have been obtained from the magnetization loops measured at constant temperatures (see Fig. 6) in $H$ oriented perpendicular ($j_c^{ab}$) and parallel ($j_c^c$) to the *ab*-plane. Arrows point maxima in the $j_c(H)$ curves.

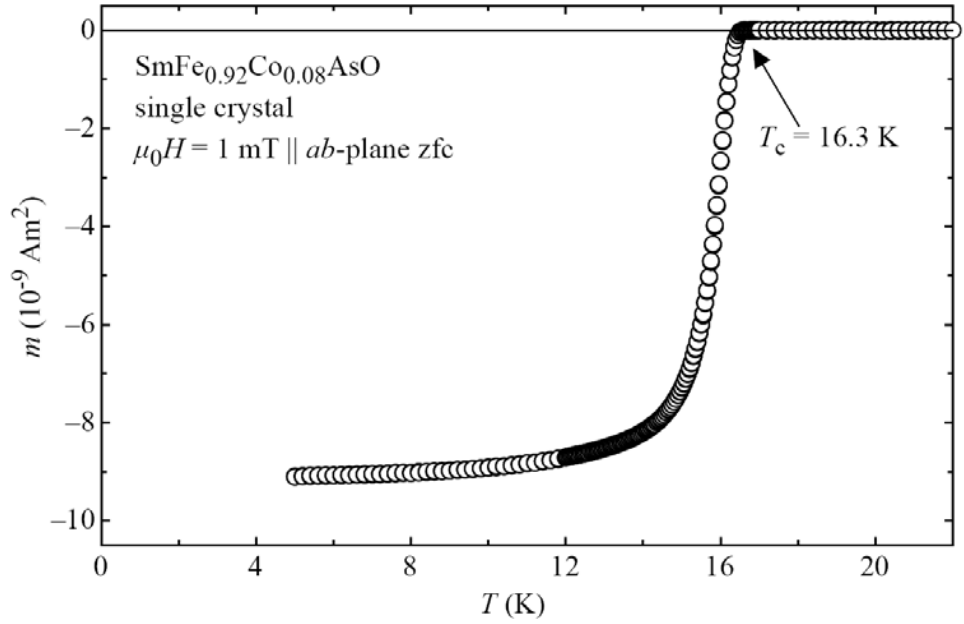

**FIG. 8.** Temperature dependence of the zero-field cooled (zfc) magnetic moment of the studied SmFe$_{0.92}$Co$_{0.08}$AsO single crystal in 1 mT magnetic field applied in its *ab*-plane. The crystal exhibits a sharp transition at $T_c$ = 16.3 K. The strong signal below $T_c$ is consistent with bulk superconductivity.

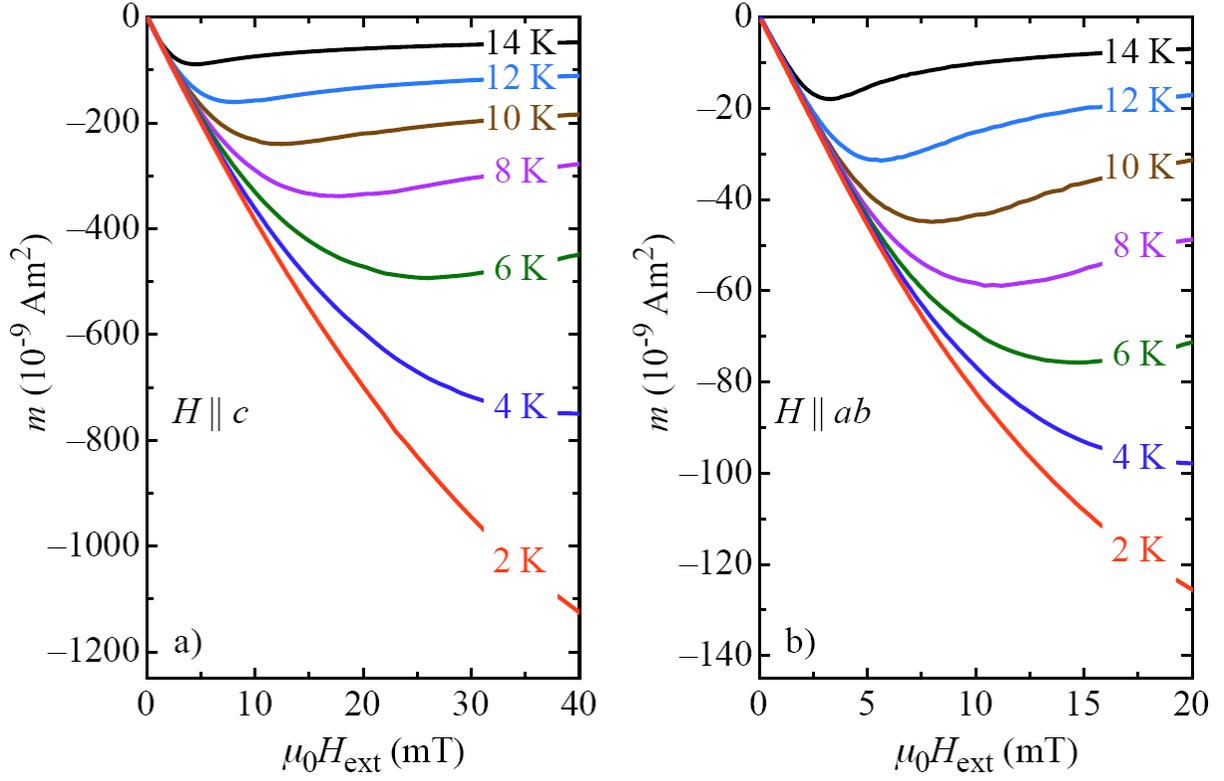

**FIG. 9.** (Color online) Magnetic moment of the studied $SmFe_{0.92}Co_{0.08}AsO$ single crystal in low external magnetic fields applied (a) along the $c$-axis and (b) in the $ab$-plane. The external field dependence of $m$ is linear in the Meissner state and show an upward curvature above $H_p$ due to the entrance of vortices into the bulk.

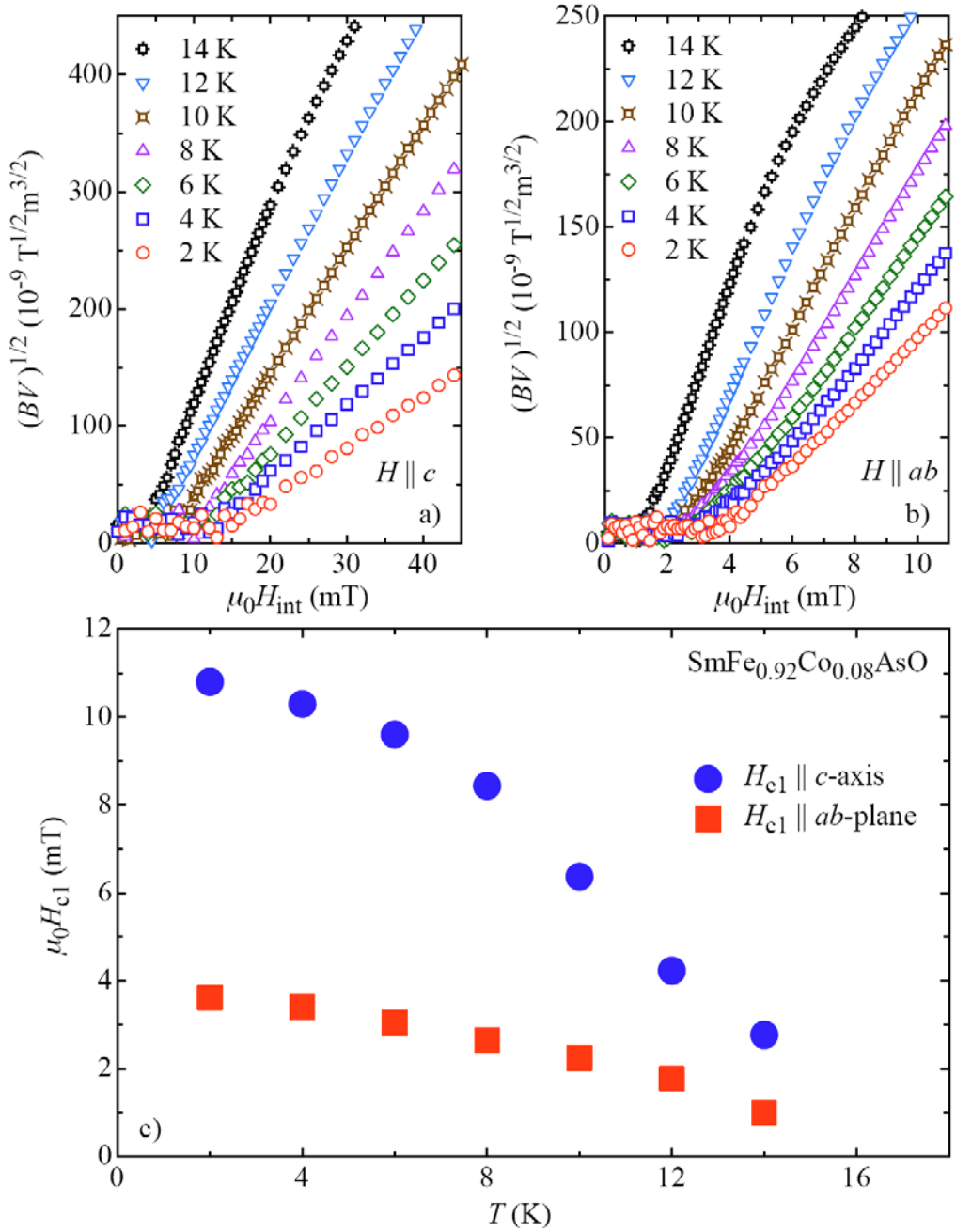

**FIG. 10.** (Color online) Analysis in order to extract the lower critical field $H_{c1}$ from $m(H)$ measurements of the SmFe$_{0.92}$Co$_{0.08}$AsO single crystal. (a), (b) Square root of $(BV)^{1/2}$ over the internal magnetic field for magnetic field along the $c$-axis and in the $ab$-plane, respectively. (c) $H_{c1}$ as a function of temperature for both crystallographic directions with $\mu_0 H_{c1}^{\|c}(0) \approx 11$ mT and $\mu_0 H_{c1}^{\|ab}(0) \approx 4$ mT.

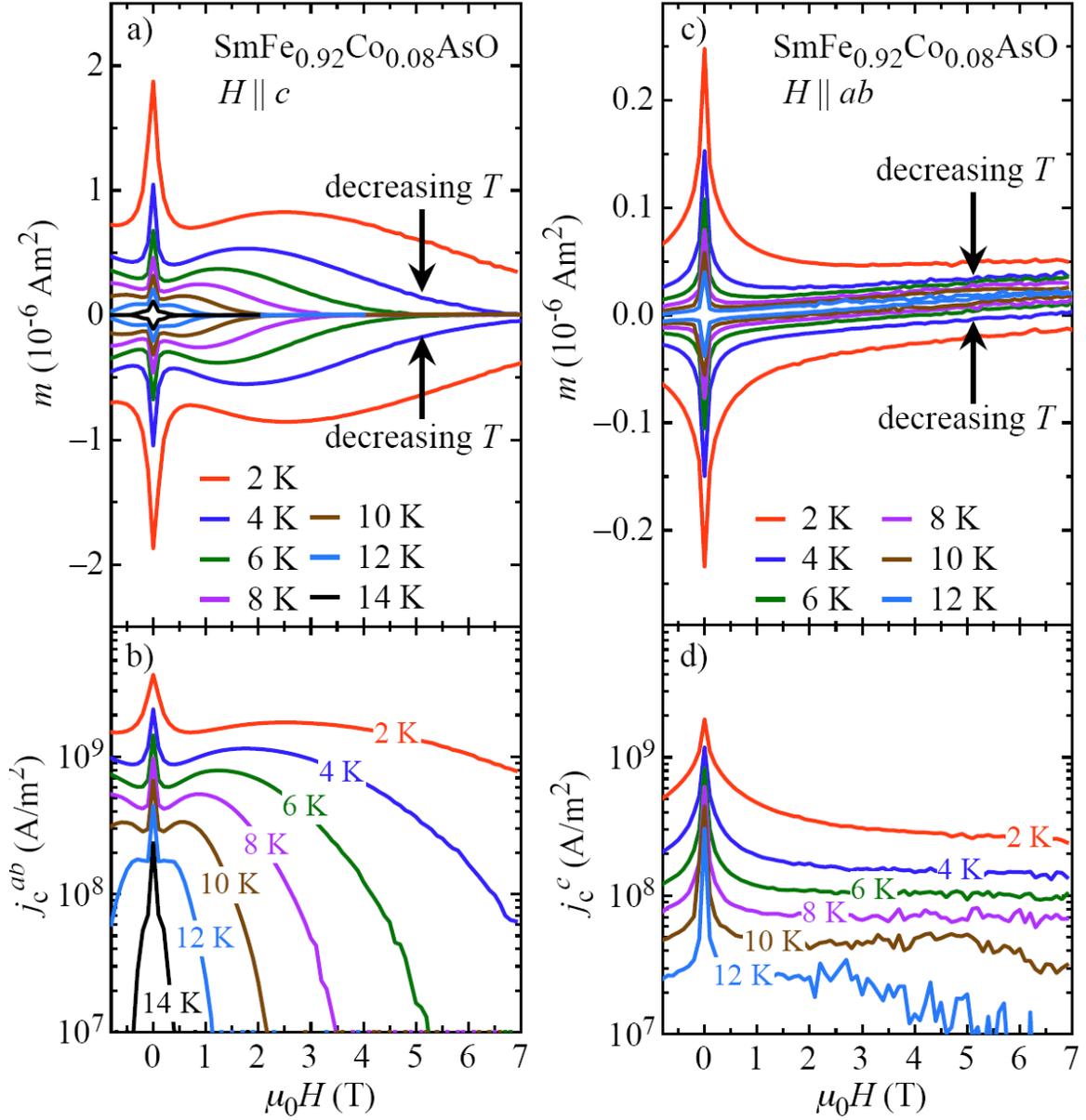

**FIG. 11.** (Color online) Critical current density in SmFe$_{0.92}$Co$_{0.08}$AsO. (a), (c) Irreversible $m(H)$ up to 7 T, measured for both, increasing and decreasing magnetic fields. (b), (d) Critical current density $j_c^c$ and $j_c^{ab}$ derived from the width of the hysteresis loops (a) and (c).

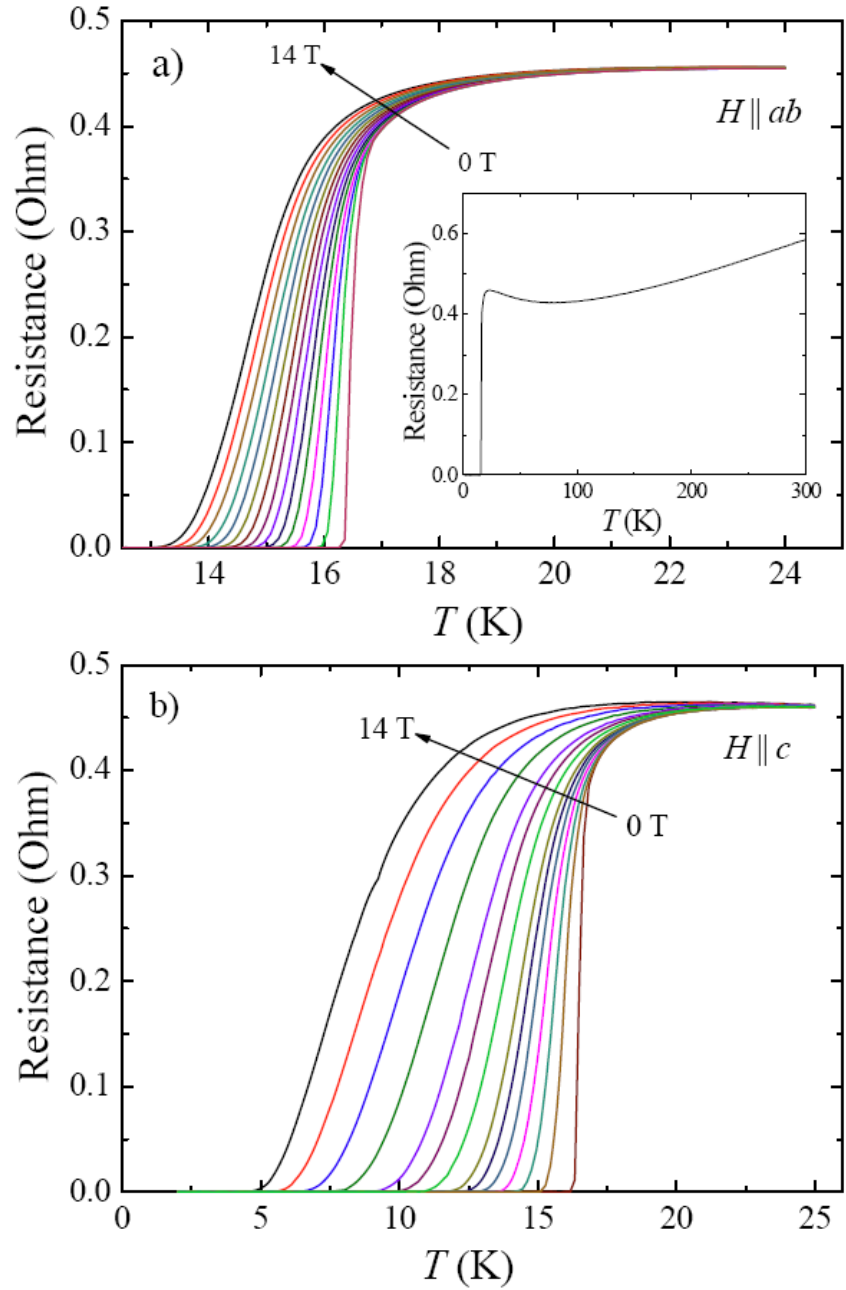

**FIG. 12.** (Color online) (a,b) Temperature dependence of the resistance for a SmFe$_{0.92}$Co$_{0.08}$AsO single crystal with the field applied along the two principal directions. Data for the magnetic field applied in the *ab*-plane were recorded in the field range 0 – 14 T with 1 T step and the data for the field along the *c*-axis were recorded in the following fields: 0, 0.5, 1, 1.5, 2, 2.5, 3, 4, 5, 6, 8, 10, 12, 14 T. Inset in (a) shows the normal state resistance.

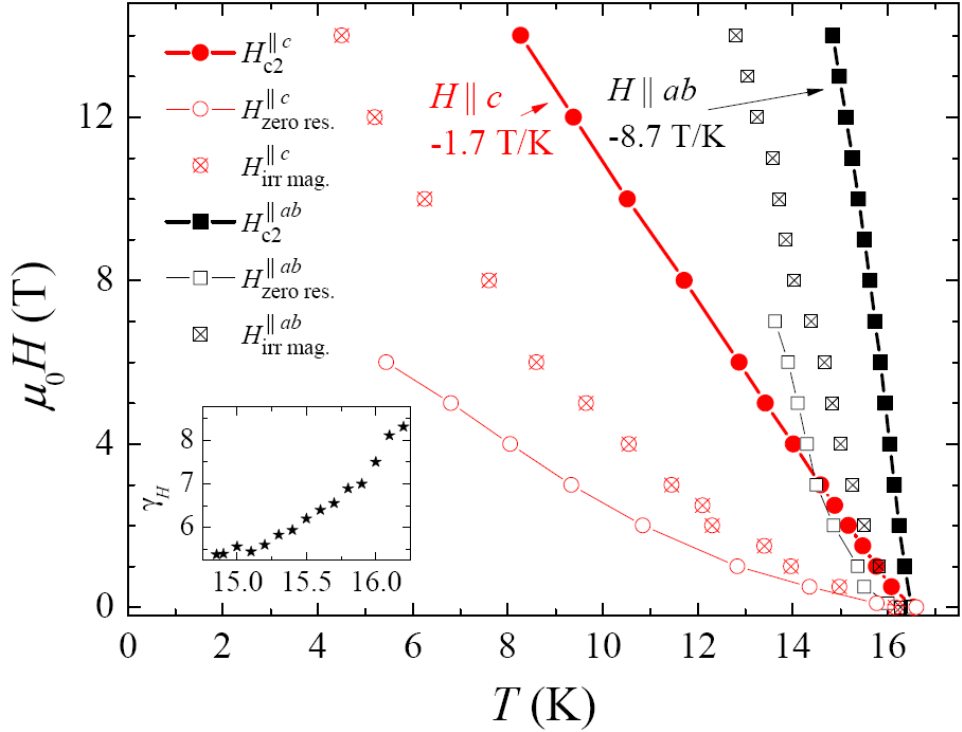

**FIG. 13.** (Color online) Phase diagram from magnetization and resistivity. The irreversibility field $H_{irr}$, estimated from the onset of irreversible magnetization in the SQUID measurements is drawn for both configurations, $H$ parallel to $c$-axis and $H$ parallel to $ab$-plane and compared with $H_{zero\ res.}$ estimated at the temperature where the onset of zero resistivity in the $\rho(T)$ recorded in magnetic field $H$ was found noticeable. The upper critical field $H_{c2}$ estimated from resistivity measurements is shown as the phase boundary. Inset: the upper critical field anisotropy $H_{c2}^{\|ab}/H_{c2}^{\|c}$ in the vicinity of $T_c$. To determine $H_{c2}$ the 50% $\rho_n$ criterion was used.

**TABLE I.** Crystallographic and structural refinement parameters of $NdFeAsO_{1-x}F_x$ and $SmFe_{0.92}Co_{0.08}AsO$ single crystals. The diffraction study was performed at 295 K using Mo Kα radiation with λ = 0.71073 Å. The lattice is tetragonal with space group *P*4/*nmm*. The absorption correction was done analytically. A full-matrix least-squares method was employed to optimize $F^2$. Some distances and marking of atoms are shown in Fig. 3.

| Sample composition | $NdFeAsO_{0.65}F_{0.35}$ | $NdFeAsO_{0.75}F_{0.25}$ | $SmFe_{0.92}Co_{0.08}AsO$ |
|---|---|---|---|
| $T_c$ (K) | 38.5 | 19 | 16.4 |
| $a$ (Å) | 3.9629(6) | 3.9643(6) | 3.9410(2) |
| $c$ (Å) | 8.5493(17) | 8.5423(14) | 8.4675(7) |
| $V$ (Å$^3$) | 134.26(4) | 134.25(4) | 131.513(14) |
| Calculated density (g/cm$^3$) | 7.198 | 7.218 | 7.510 |
| $z_{Ln}$ | 0.1421(1) | 0.1414(2) | 0.1374(1) |
| $z_{As}$ | 0.6583(2) | 0.6587(4) | 0.6593(1) |
| $Ln_1$-$Ln_2$ (Å) | 3.7093(15) | 3.700(2) | 3.6307(3) |
| O-O= Fe-Fe (Å) | 2.8022(3) | 2.8032(3) | 2.78671(14) |
| $Ln_2$-$As_1$ (Å) | 3.2809(13) | 3.282(2) | 3.277(2) |
| $Ln$-O (Å) | 2.3244(6) | 2.321(1) | 2.28841(15) |
| $As_1$-$As_2$ (Å) | 3.8660(17) | 3.8999(34) | 3.8784(6) |
| Fe-As (Å) | 2.3993(12) | 2.401(2) | 2.3879(3) |
| $As_1$-Fe-$As_2$, $\beta$ (deg) | 108.54(4) | 108.58(7) | 108.60(1) |
| $As_2$-Fe-$As_3$, $\alpha$ (deg) | 111.35(8) | 111.26(14) | 111.22(2) |
| $S_3^a$ (Å) | 1.706(2) | 1.708(5) | 1.721(1) |
| $S_1^a$ (Å) | 2.430(2) | 2.416(5) | 2.327(2) |
| $h_{Pn}^a$ ($S_2/2$) (Å) | 1.353(2) | 1.356(4) | 1.3489(9) |
| $h_{Ln}^a$ ($S_1/2$) (Å) | 1.215(1) | 1.208(3) | 1.163(6) |
| Absorption coefficient (mm$^{-1}$) | 36.410 | 36.421 | 39.82 |
| $F(000)$ | 260 | 254 | 258 |
| Crystal size, (μm$^3$) | 56 × 272 × 426 | 25 × 95 × 174 | 38 × 126 × 129 |
| $\theta$ range for data collection | 2.38° - 43.47° | 2.38° - 43.08° | 4.81° - 50.01° |
| Index ranges | -3≤h≤7, -7≤k≤7, -7≤l≤14 | -6≤h≤7, -7≤k≤5, -14≤l≤16 | -7≤h≤5, -8≤k≤8, -17≤l≤14 |
| Reflections collected/unique | 1052/332 $R_{int.}$= 0.0558 | 1908/329 $R_{int.}$= 0.0780 | 2587/441 $R_{int.}$= 0.0472 |
| Completeness to $2\theta$ | 94.9 % | 96.2 % | 96.3 % |
| Data/restraints/parameters | 332/0/11 | 329/0/11 | 441/0/11 |
| Goodness of fit on $F^2$ | 1.335 | 1.376 | 1.154 |
| Final $R$ indices [$I$>2σ($I$)] | $R_1$ = 0.0513, w$R_2$ = 0.1544 | $R_1$ = 0.0901, w$R_2$ = 0.1897 | $R_1$ = 0.0241, w$R_2$ = 0.0572 |
| $R$ indices (all data) | $R_1$ = 0.0537, w$R_2$ = 0.1553 | $R_1$ = 0.1017, w$R_2$ = 0.1953 | $R_1$ = 0.0245, w$R_2$ = 0.0573 |

[a]Fig. 3